\documentclass[twocolumn,showpacs,preprintnumbers,amsmath,amssymb,pra]{revtex4-1}
\usepackage{epsfig} %Works faster than the graphicx package}
\usepackage{dcolumn}% Align table columns on decimal point
\usepackage{bm}% bold math
\usepackage{epstopdf}
\begin{document}

% \preprint{APS/123-QED}

\title{Measurement of the Stark shift of the $6s^2S_{1/2} \rightarrow 7p^2P_{J} $ transitions in atomic cesium  }

\author{George Toh$^1$, D. Antypas$^{1,2}$, and D. S. Elliott$^{1,2}$}
\affiliation{%
   $^1$School of Electrical and Computer Engineering and $^2$Department of Physics and Astronomy\\ Purdue University, West Lafayette, IN  47907
}

\date{\today}% It is always \today, today,

\begin{abstract}
We report measurements of the Stark shift of the cesium $6s \: ^2S_{3/2} \rightarrow 7p \: ^2P_{3/2} $ and the $6s \: ^2S_{1/2} \rightarrow 7p \: ^2P_{1/2} $ transitions at $\lambda = 456$ nm and 459 nm, respectively, in an atomic beam.  From these, we determine the static scalar polarizability for both 7P states, and the tensor polarizability for the 7P$_{3/2}$ state.  The fractional uncertainty of the scalar polarizabilites is $\sim$0.18\%, while that of the tensor term is 0.66\%. These measurements provide sensitive tests of theoretical models of the Cs atom, which has played a central role in parity nonconservation measurements.
\end{abstract}

\maketitle

\section{Introduction}
The high precision attainable in measurements of the Stark shift of atomic transition frequencies makes them sensitive tests of theoretically determined radial matrix elements.  Atomic cesium, which has played a central role in parity nonconservation measurements over the past forty years~\cite{BouchiatB75,WoodBCMRTW97,PorsevBD09,JohnsonSS03,LintzGB07,RobertsDF13a}, is of particular interest in this regard, where accurate determinations of electric dipole matrix elements, experimental~\cite{YoungHSPTWL94,RafacT98,RafacTLB99,AminiG03,DereviankoP02,BouloufaCD07,AntypasE13c,SieradzanHS04,VasilyevSSB02} and theoretical~\cite{PorsevBD10,IskrenovaSS07,SafronovaC04}, are critical for precise determination of the weak charge.  In this report, we discuss our recent measurements of the scalar static polarizabilities $\alpha_0$ of the $7p ^2P_{1/2}$ and $7p ^2P_{3/2}$ states of atomic cesium, whose magnitudes depend primarily upon the reduced terms $\langle 7P_{1/2} || r || 6D_{3/2} \rangle$ and $\langle 7P_{3/2} || r || 6D_{5/2} \rangle$, respectively, and the tensor static polarizability $\alpha_2$ of the $7p ^2P_{3/2}$ states of atomic cesium, whose magnitude contains primary contributions from the reduced terms $\langle 7P_{3/2} || r || 6D_{5/2} \rangle$, $\langle 7P_{3/2} || r || 6D_{3/2} \rangle$, $\langle 7P_{3/2} || r || 8S_{1/2} \rangle$, and $\langle 7P_{3/2} || r || 7S_{1/2} \rangle$~\cite{IskrenovaSS07}.  

\begin{figure}
	  \includegraphics[width=8cm]{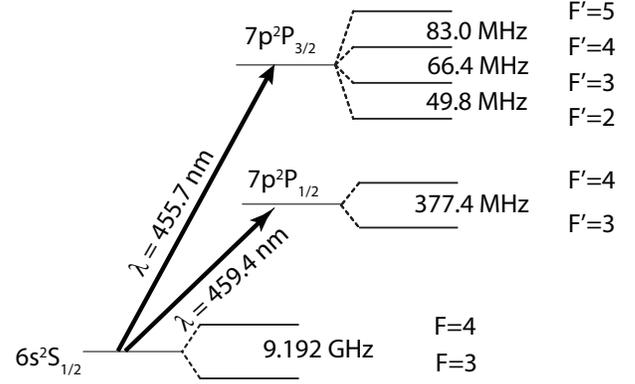}\\
	  \caption{Energy level diagram of atomic cesium, showing the levels relevant to these measurements.  Hyperfine splittings of the 7P states are taken from Ref.~\protect\cite{ArimondoIV77}}
	  \label{fig:EnergyLevels}
	\end{figure}

The tensor polarizability for the $7p ^2P_{3/2}$ state in cesium has been measured previously using the level-crossing technique by Khadjavi, Lurio, and Happer~\cite{KhadjaviLH68} and by Khvostenko and Chaika~\cite{KhvostenkoC68}, with a measurement uncertainty of a few percent in each case.  A subsequent measurement of this tensor polarizability, as well as the scalar polarizability $\alpha_0$ for both $7p ^2P_{J}$ lines, was reported by Domelunksen~\cite{Domelunksen83}, with a comparable uncertainty.  In the present work, we are able to improve the precision of each of these polarizabilites.  To achieve this, we use narrow-band, frequency-stabilized diode lasers to excite the $6s ^2S_{1/2} \rightarrow 7p ^2P_{J}$ transitions in a nearly Doppler-free atomic beam geometry, allowing us to spectrally resolve the various hyperfine components of the transitions (shown schematically in Fig.~\ref{fig:EnergyLevels}).  We report values of $\alpha_0$ with an uncertainty of $\sim$0.18\% and of $\alpha_2$ with an uncertainty of 0.66\%.  Our results are in good agreement with early theoretical values based upon Coulomb potentials~\cite{WijngaardenL94}, as well as the more recent results of Iskrenova-Tchoukova, Safronova, and Safronova~\cite{IskrenovaSS07}, who use a relativistic, all-order method to calculate transition moments, and a sum-over-states method to determine the polarizabilities.  

\begin{figure}
	  \includegraphics[width=9cm]{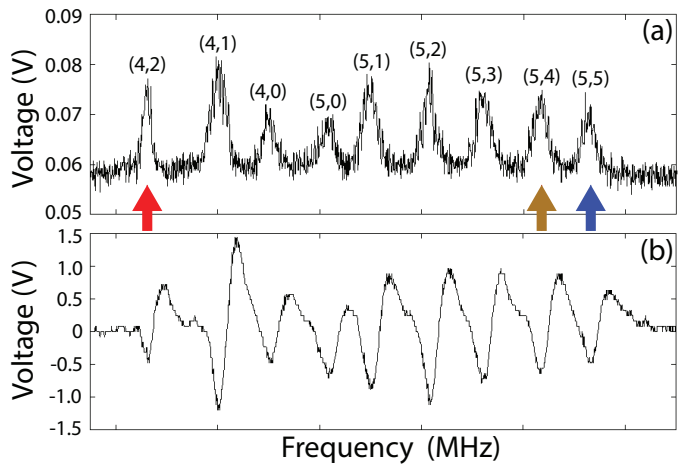}\\
	  \caption{(Color on-line) (a) An uncalibrated partial Stark spectrum of the $6s ^2S_{1/2}, F=4 \rightarrow 7p ^2P_{3/2}$ transition at $E_0 = 6$ kV/cm showing the splitting of the lines.  The notation above each peak indicates $(F^{\prime}, \: |m_F^{\prime}| )$ of the $7p ^2P_{3/2}$ state. The arrows indicate the peaks used in our measurements, and the frequency difference between the red and blue arrows is approximately $155$ MHz.  (b) The derivative signal of the same spectrum. There is a small mis-match between the error signal zero crossing and the center of the signal peaks due to the long integration time constant on our lock-in amplifier. This does not affect our results because we fix the EOM frequency when making a measurement.} %mf=5 and mf=4 peaks are ~17 MHz apart
	  \label{fig:6kVspectrumwithES}
	\end{figure}

Upon application of a dc electric field of magnitude $E_0$ to an atomic system, the energy of a bound state of that atom is shifted through the quadratic Stark effect by an amount
	\begin{equation}
	\Delta E = -\frac{1}{2} \alpha E_0^2,
	\end{equation}
where $\alpha$ is the polarizability of the atomic level.  For a level of electronic angular momentum $J$, the polarizability $\alpha$ can be expressed in terms of its scalar ($\alpha_0$) and tensor ($\alpha_2$) components as 
	\begin{equation}\label{eq:alphaJscalten}
	\alpha = \alpha_0 + \alpha_2 \: \frac{3 m_J^2 - J(J+1)}{J (2J-1)} ,
	\end{equation}
where $m_J$ is the projection of the angular momentum on the $z$ axis.  The scalar term represents an overall shift of all components of the level together, while $\alpha_2$ describes a splitting of the state into its various magnetic components.  For $J=0$ or $\frac{1}{2}$, the $\alpha_2$ term in Eq.~(\ref{eq:alphaJscalten}) vanishes, and the level is shifted in energy, but remains unsplit. For levels that exhibit hyperfine structure and have angular momentum $J \ge 1$, the Stark effect produces a much richer spectrum.  This has been described by Schmieder~\cite{Schmieder72} through a polarizability $\alpha$ of the form
	\begin{equation}\label{eq:polQ}
	\alpha = \alpha_0 + \alpha_2 \: Q_{F,\tilde{F}; |m_F|},
	\end{equation}
where the matrix $Q_{F,\tilde{F}; |m_F|}$ describes the mixing of states of unequal $F$ ($F$ and $\tilde{F}$ in this expression), but equal $m_F$, by the static field.  (We use the usual notation here, with $F$ representing the total atomic angular momentum, including nuclear spin $I$, and $m_F$ the projection of $F$ on the $z$ axis.)  The scalar part of the polarizability shifts all hyperfine and magnetic sublevels equally, while the tensor part causes the spectrum to diverge into a series of individual lines.  As an illustration, we show an uncalibrated, partial Stark spectrum of the $6s ^2S_{1/2} \rightarrow 7p ^2P_{3/2}$ transition at $E_0 = 12$ kV/cm in Fig.~\ref{fig:6kVspectrumwithES}(a). 
We label each peak with $F^{\prime}$ and $m_F^{\prime} $ of the $7p ^2P_{3/2}$ state.  In contrast, each hyperfine line of the $6s ^2S_{1/2} \rightarrow 7p ^2P_{1/2}$, while shifted by the Stark effect, remains a single line. In the following, we will discuss our experimental observations of the Stark spectrum of the two transitions $6s ^2S_{1/2} \rightarrow 7p ^2P_{J}$, J = $\frac{1}{2}$ and $\frac{3}{2}$, in atomic cesium, and from these  our determination of the scalar and tensor polarizabilities.  

\section{Description of apparatus}\label{sec:apparatus}
\begin{figure}
	  \includegraphics[width=8cm]{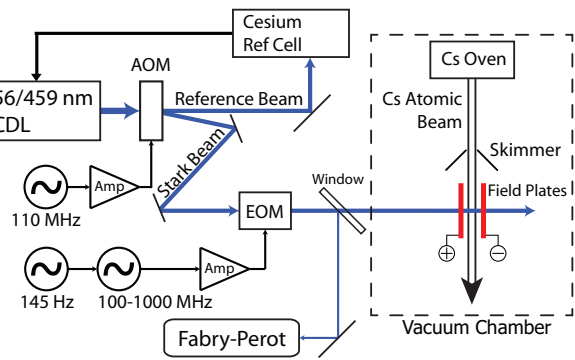}\\
	  \caption{(Color on-line) Diagram of the experimental configuration.  The external cavity diode laser (ECDL) generates light at 455.7 nm or 459.4 nm, resonant with the $6s ^2S_{1/2} \rightarrow 7p ^2P_{3/2}$ or $6s ^2S_{1/2} \rightarrow 7p ^2P_{1/2}$ transition, respectively.  The frequency of this beam is shifted in the acousto-optic modulator (AOM), and sidebands imposed in the electro-optic modulator (EOM), before crossing the atomic beam inside the vacuum chamber at a perpendicular crossing angle. We monitor the sideband structure of the Stark beam with the Fabry-Perot interferometer.  }
	  \label{fig:Experimentalsetup}
	\end{figure}

The general principle of the measurement is similar to that of several  other recent works~\cite{TannerW88, WijngaardenHLR94, AntypasE11, KortynaTGSS11}.  We use the output of a single, narrow-band tunable laser source, which we split into two separate beams, labeled the reference and Stark beams in Fig.~\ref{fig:Experimentalsetup}.  Using an electro-optic modulator (EOM) and an acousto-optic modulator (AOM) to offset the frequencies of these two beams, we concurrently bring the reference beam into resonance with the cesium transition in a field-free vapor cell (the reference cell), and the Stark beam into resonance with the transition in cesium atoms to which a uniform electric field has been applied.  The difference between the frequency offsets of these two beams, which depends only on the rf frequencies driving the modulators, equals the Stark-shift of the resonance.  This eliminates the requirement for calibration of the laser frequency scan, which can be problematic at the precision required in these measurements.  The Doppler broadening of the resonances is largely suppressed in our measurements, allowing us to resolve the hyperfine structure of the transitions, and also allowing us to use relatively low dc electric field strengths in our measurements.
 
The laser for these measurements, which we operate at wavelengths of 455.7 (for the $6s ^2S_{1/2} \rightarrow 7p ^2P_{3/2}$ transition) or 459.4 nm (for the $6s ^2S_{1/2} \rightarrow 7p ^2P_{1/2}$ transition), is a home-made external cavity diode laser (ECDL) using an AR coated laser diode, which generates approximately 10 mW of optical power.  
We diffract the output beam in an AOM, and use the first-order diffracted beam, whose frequency is $f_l + f_{AO}$ (where $f_l$ is the frequency of the laser output and $f_{AO}$ = 110.0 MHz is the AOM drive frequency), for the experiment (i.e. this is the Stark beam).  The 110 MHz drive signal is produced by a synthesized signal generator and amplified by an rf amplifier.  We direct the undiffracted beam, which we use as our reference beam, into a field-free cesium vapor cell, and frequency-lock the laser to one hyperfine component of the Doppler-free saturated absorption spectrum (the $6s ^2S_{1/2}, \: F=4 \rightarrow 7p ^2P_{1/2}, \: F=4$ line at $\lambda$ = 459 nm or the $6s ^2S_{1/2}, \: F=4 \rightarrow 7p ^2P_{3/2}, \: F=5 $ line at $\lambda$ = 456 nm) of this spectrum.  To obtain an error signal for locking to the peak of the hyperfine line, we dither the laser injection current at 30 kHz.  
In either case, the laser frequency $f_l$ is resonant with and stabilized to the unshifted atomic resonance, $f_a$.
The linewidth of the laser spectrum is $<1$ MHz. Because the absorption strengths of these transitions are relatively weak, we have to heat the cesium vapor cell to a temperature in the range 80-110$^{\circ}$C to obtain sufficient Cs density within the cell. 

We impose optical sidebands on the spectrum of the Stark beam by modulating its phase in a traveling wave EOM, driven by a separate signal generator and amplifier at a frequency $f_{EO}$, where we can adjust $f_{EO}$ to any frequency in the range from 110 to 1000 MHz.  
We use the lower frequency sideband, whose frequency is $f_{l} + f_{AO} - f_{EO}$, to excite the Stark-shifted absorption resonance in the atomic beam.  We carry out the measurements in one of two different modes: we fix the frequency $f_{EO}$ and vary the dc field amplitude $E_0$ to `tune' the Stark-shifted absorption line into resonance with the lower frequency sideband; or we fix the amplitude $E_0$ and vary the frequency $f_{EO}$ to match the Stark-shifted resonance.

The cesium atom beam is formed inside an aluminum vacuum chamber pumped with a turbomolecular pump to a pressure of $5 \times 10^{-6}$ torr.  We use an effusive cesium oven with a nozzle consisting of an array of stainless steel hypodermic needle tubes to form the atom beam. More details are available in our earlier publications~\cite{AntypasE13a, AntypasE13b}. This oven and nozzle generates a beam of dimension 12 mm $\times$ 8 mm near the nozzle.
We insert an atomic beam aperture (labeled skimmer in Fig.~\ref{fig:Experimentalsetup}) before the interaction region to further reduce the width of the atomic beam to $\sim\frac{1}{2}$ the spacing of the field plates. This reduces the accumulation of cesium on the electric field plates.  The spectral width of the $6s \rightarrow7p$ absorption lines in our beam geometry is $\sim$6 MHz FWHM, largely due to Doppler broadening in the slightly diverging atomic beam.  The natural linewidths of these transitions, corresponding to the 133 ns lifetime of $7P_{3/2}$ and the 155 ns lifetime of the $7P_{1/2}$ state~\cite{PaceA75,CampaniDG78,OrtizC81}, are 1.2 MHz and 1.0 MHz, respectively. 

The uniformity of the static electric field, and the precision with which this field can be determined, depends critically on the parallel conducting field  plates used to generate this field.  We construct these field plates from a pair of $76.2 \times 25.4$ mm ($3 \times 1$ in) microscope glass slides, coated on the inside surfaces with a thin conducting layer of indium tin oxide (ITO). These field plates are spaced by $4.928 ~(4)$ mm ($0.19400 ~(15)$ in), and are mounted inside an aluminum framework with external ceramic posts using a vacuum compatible epoxy. (The number enclosed within parentheses following these parameters indicates our estimate of the uncertainty.)
We evaluated the non-uniformity of the electric field within the interaction region due to fringing effects using a commercial software package, and found that this variation is less than a part in $10^5$.

During assembly, we spaced the field plates with a set of carefully selected ceramic spacers to assure a high degree of parallelism, then removed the spacers after the epoxy had dried. (We observed drifts in some of our early Stark shift measurements, which we attributed to an accumulation of cesium on the internal spacers used for those measurements. These drifts were absent after we removed the internal spacers.) We estimate the $0.00015$ inch uncertainty in the spacing of the glass slides based on the relative ease with which we can slip calibrated ceramic beads, whose lengths we measured to $\pm$0.00005 inch, at various locations near the central region of the field plates, similar to the technique described in Refs.~\cite{WijngaardenHLR94,AntypasE11}. We also measured the parallelism of the plates by reflecting a HeNe laser beam from the two surfaces, and observing the spacing of the fringes formed by the interference of the two reflected beams. We estimate that the angle between the two plates was less than 0.15 mrad. This high degree of parallelism between the field plates is consistent with our estimate of the variation of the plate spacing over the width of the plates.

We use a pair of stable high-voltage sources to bias the field plates, one plate positively biased, the other negative. Between sets of data, we switch the polarity of the field plates.  We measure the voltage applied to each field plate using an Ohmcraft 1000:1 high voltage resistive divider, which we have carefully checked and calibrated for nonlinearity and stability.  The fractional uncertainty in the voltage measurement of each field plate is $\sim 2 \times 10^{-5}$.

Consistent with the treatment by Schmieder~\cite{Schmieder72}, we define the $z$-axis of the atomic system as the direction of the applied field $E_0$. While the Stark beam for these measurements propagates in a direction $\hat{\mathbf k}$ nearly parallel to this $z$-axis, and its polarization state is linear, the experiment is relatively insensitive to either of these conditions, since the ground state components are degenerate, and the various peaks in the Stark spectrum correspond to different hyperfine components of the excited state alone. Changes in polarization or imperfect alignment of $\hat{\mathbf k}$ with the $z$-axis only change the relative height of the peaks in the Stark spectrum, but not their frequency.  By contrast, it is important that the laser beam propagates in a direction perpendicular to the atomic velocity to assure narrow absorption linewidths and to minimize the Doppler shift of the lines.
Using an alignment laser, we mount the parallel field plates inside the vacuum system centered on and parallel to the atomic beam.  In addition, we observe the reflection of the Stark beam from the field plates and adjust this beam to normal incidence on the field plates.  After these alignment steps, only a minor adjustment of the Stark beam direction $\hat{\mathbf k}$ is necessary to minimize the Doppler shift of the resonance in the atomic beam, which we determine by zeroing the applied field and comparing the absorption resonance in the atomic beam to that of the reference cell.

In order to detect the absorption resonances in the atomic beam, we use the detection system that we developed earlier~\cite{AntypasE13a, AntypasE13b} for sensitive measurement of highly-forbidden optical transitions.  We based this system on a technique reported earlier in Ref.~\cite{WoodBCMRTW97}. The population of the cesium atoms as they effuse from the oven is equally distributed among each of the  F=3 and F=4 hyperfine components of the ground state.  Before the atoms interact with the blue laser, we transfer all of the atoms into the F=4 level by optically pumping the population with the output of an 852 nm ECDL tuned to the $6s ^2S_{1/2}, \: F=3 \rightarrow 6p ^2P_{3/2}, \: F^{\prime}=4 $ hyperfine transition of the D$_2$ resonance line.  
After interacting with the blue laser, the population in the $F=3$ ground state is a measure of the excitation rate by the Stark beam to the $7P$ state, since these atoms decay spontaneously back to the ground state, where some end up in the initially empty F=3 hyperfine state. We detect this population using the output of a second ECDL tuned to the D$_2$ line at 852 nm (in this case resonant with the $6s ^2S_{1/2}, \: F=3 \rightarrow 6p ^2P_{3/2}, \: F^{\prime}=2 $ cycling transition) and a large area photodiode to measure the scattered optical power in this region.

We use a lock-in amplifier for phase-sensitive detection of the photodiode current in order to improve the sensitivity of the measurement.  We dither the  frequency of the EO sideband at 145 Hz (with a 1 MHz amplitude), which modulates the rate of absorption by the atoms. The derivative signal produced by the lock-in amplifier, illustrated in Fig.~\ref{fig:6kVspectrumwithES}(b), is a dispersion shaped resonance of width $\sim$6 MHz.  The zero-crossing is well suited for determination of line center.

During the course of our measurements, we found that the amplitude of the optical sideband of the Stark beam, as monitored with a scanning Fabry-Perot interferometer, varied across the 100-1000 MHz spectrum. To ensure that the optical power in the sideband used for the experiment is constant, we selected frequencies $f_{EO}$ at which the sideband power is relatively uniform. 

We must keep the laser intensity below the saturation intensity $I_{\rm sat}$ in order to minimize power broadening and light shifts. The laser intensity for these measurements is $4$ mW/cm$^2$, of which only $\sim \frac{1}{3}$ is in the lower sideband that interacts with the atoms.  Using the reduced matrix dipole matrix elements for these transitions~\cite{AntypasE13c}, we estimate that the saturation intensity $I_{\rm sat}$ of the $6s ^2S_{1/2} \rightarrow 7p ^2P_{3/2}$ transition is about 15 mW/cm$^2$, while that of the $6s ^2S_{1/2} \rightarrow 7p ^2P_{1/2}$ line is about 50 mW/cm$^2$. Therefore, the sideband intensity is well below $I_{\rm sat}$ in both cases.

\subsection{ Scalar polarizability of $7p ^2P_{1/2}$}
\begin{figure}
	  \includegraphics[width=8cm]{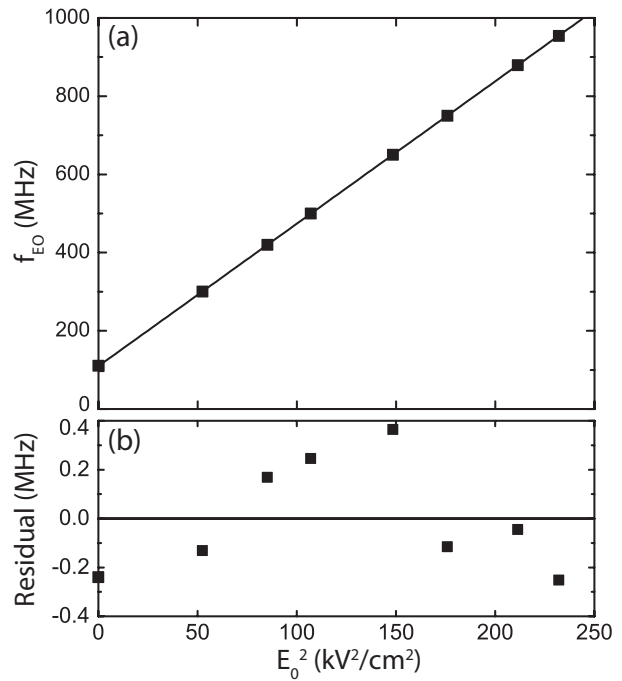}\\
	  \caption{(a) A plot of $f_{EO}$ vs $E_0^2$ for the $6s ^2S_{1/2} \rightarrow 7p ^2P_{1/2}$ transition.  Experimental data is shown by the square points, while the solid line shows the result of a linear least squares fit. The residual error of each data point is shown in (b). }
	  \label{fig:p12poscorrStraightfitCombined}
	\end{figure}

For our determination of the polarizability $\alpha_0$ of the $7p ^2P_{1/2}$ state, we first set the EO modulation frequency $f_{EO}$ to one of seven pre-determined values in the range between 110 and 1000 MHz.  (The minimum of this range is the AO drive frequency, and corresponds to a zero Stark shift, while the maximum frequency is the maximum frequency of our signal generator.)  At each value of $f_{EO}$, we adjust the voltage applied to the plates to shift the transition into resonance with the lower sideband of the Stark beam.  In Fig.~\ref{fig:p12poscorrStraightfitCombined}, we show one set of these data, plotted as $f_{EO}$ vs. $E_0^2$.  The solid line indicates the result of a linear least squares fit with two adjustable parameters, the intercept and the slope. 

The intercept of this fitted line is 109.8 (2) MHz, consistent with the 110.0 MHz frequency offset imposed by the AOM.
The slope of this line is 3.6417 (12) MHz/(kV/cm)$^2$, and is equal to half the difference between the polarizabilities of the $7P_{1/2}$ state and the $6S_{1/2}$ state, $\frac{1}{2}\{\alpha_0(7P_{1/2}) - \alpha_0(6S_{1/2})\}$. The uncertainty of $0.0012$ MHz/(kV/cm)$^2$ is statistical and is determined from the scatter of the data points from the linear fit to the data.  We show the difference between the data points and the linear fit, in Fig.~\ref{fig:p12poscorrStraightfitCombined}(b).  The rms residual for this set of data is 0.22 MHz. 
  
\begin{figure}
	  \includegraphics[width=8cm]{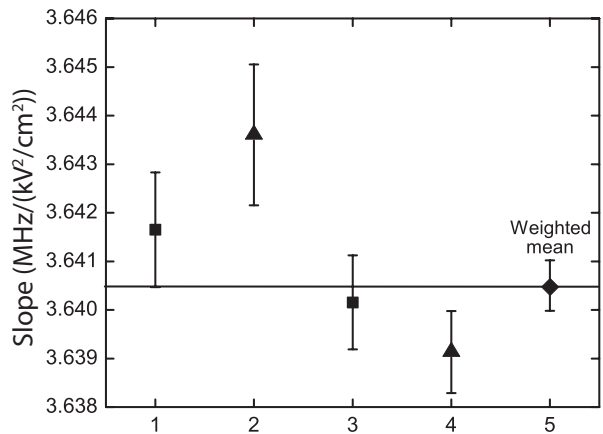}\\
	  \caption{Summary of slopes for the 4 measurements made. Error bars shown are statistical only. Square data points were obtained with electric field in the same direction as the Stark beam. For the triangular data points, the measurements were made with the electric field orientation reversed. The fifth point (diamond-shaped) and horizontal line denote the weighted average.}
	  \label{fig:p12summary}
	\end{figure}

We measured the Stark shift of the $7P_{1/2}$ state four times, reversing the direction of electric field between sets of data and observed no systematics due to electric field orientation. We show the slope $\frac{1}{2}\{\alpha_0(7P_{1/2}) - \alpha_0(6S_{1/2})\}$ resulting from each of these measurements in Fig.~\ref{fig:p12summary}. The error bars shown in the figure indicate the statistical uncertainty for each data point.
	
The reduced $\chi^2$ for these measurements is 4.14, indicating that the measurement uncertainty is larger than the statistical uncertainty. The weighted average of the four measurements yields $\frac{1}{2}\{ \alpha_0(7P_{1/2}) - \alpha_0(6S_{1/2}) \} = 3.6405 \: (6)$ MHz/(kV/cm)$^2$, as indicated by the diamond-shaped data point and horizontal line in Fig.~\ref{fig:p12summary}. We have not scaled the statistical error despite the the large $\chi^2$ factor. As we will discuss in Section~\ref{sec:errors}, the overall measurement uncertainty is dominated by the uncertainty in the field plate spacing, and scaling the statistical error has little impact on our final result. 

Using the ground state polarizability $\alpha_0(6S_{1/2}) = 0.09978 \: (7)$ MHz/(kV/cm)$^2$ from Ref.~\cite{AminiG03}, we find $\alpha_{0} = 7.3808 \: (12)$ MHz/(kV/cm)$^2$, where the number in parenthesis denotes the statistical error only. In atomic units, this converts to $29,662 \: (5) \: a_o^3$.

\subsection{Scalar and tensor polarizability of $7p ^2P_{3/2}$}

We base our determinations of $\alpha_0$ and $\alpha_2$ for the $7p ^2P_{3/2}$ line on two lines in the Stark-shifted spectrum, namely the ($F^{\prime}$, $|m_F^{\prime}|$) = (5,5) line and the (4,2) line. We chose these particular $m_F$ peaks because they are well resolved from other peaks in the spectrum, as shown in Fig.~\ref{fig:6kVspectrumwithES}, and because their frequency difference due to the Stark shift is large, allowing for a more precise evaluation of $\alpha_2$. 
From Eq.~(\ref{eq:polQ}), the polarizability of the $F^{\prime}=5$, $m_F=\pm5$ components is $\alpha_0 + \alpha_2$. Our process for determination of the sum $\alpha_0  + \alpha_2$ for the $7p ^2P_{3/2}$ state then is similar to that of $\alpha_0$ for the $7p ^2P_{1/2}$, described in the last section.  With the reference laser frequency $f_l$ tuned and locked to the $6s ^2S_{1/2}, \: F=4 \rightarrow 7p ^2P_{3/2}, \: F^{\prime} =5$ resonance in the reference cell, we adjust the frequency $f_{EO}$ of the signal applied to the EOM to one of seven values in the range from 700 to 1000 MHz.  (Below 700 MHz, the various peaks within the Stark spectrum partially overlap, introducing errors in the measurements of the line center.)  Then we vary the voltage applied to the field plates to bring the (5,5) peak into resonance. We also take one measurement at zero electric field, varying $f_{EO}$ to find the line center.

	\begin{figure}
	  \includegraphics[width=8cm]{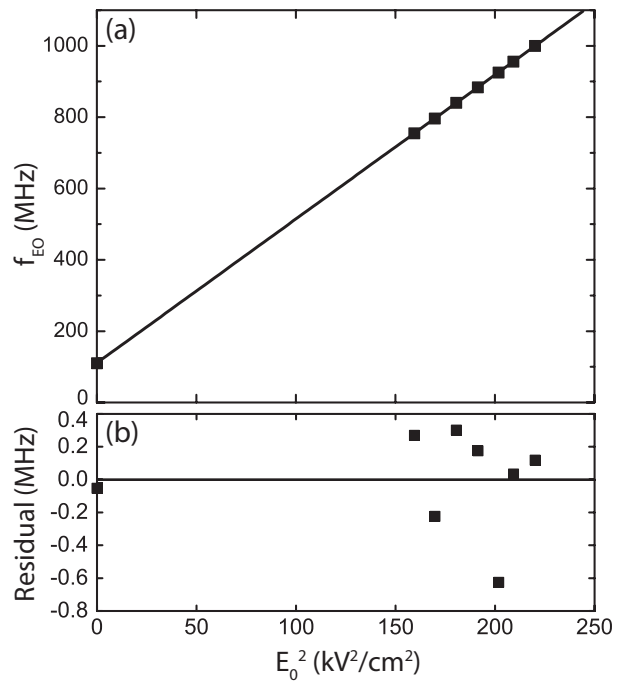}\\
	  \caption{(a) An example of one data set of $f_{EO}$ vs. $E_0^2$ for the $6s ^2S_{1/2}, \: F=4, \: m_F=\pm 4 \rightarrow 7p ^2P_{3/2}, \: F^{\prime} =5, \: m_F^{\prime}=\pm 5 $.  The straight line is the result of a linear least squares fit to the data points.  (b) The residual between the data points and the straight line. }
	  \label{fig:p32posStraightfitCombined}
	\end{figure}

We show an example of one data set in Fig.~\ref{fig:p32posStraightfitCombined}.	
The result of a linear least squares fit, represented by the straight line in this figure, yields an intercept of 110.7 (3) MHz and a slope of $\frac{1}{2}\{ \alpha_0 (7P_{3/2}) + \alpha_2(7P_{3/2}) - \alpha_0 (6S_{1/2}) \} =  4.0386 \: (18)$ MHz/(kV/cm)$^2$. We show the deviation of each of the data points from the fitted line in Fig.~\ref{fig:p32posStraightfitCombined}(b).  The rms residual is 0.3 MHz. 
We repeat this measurement with the electric field orientation reversed, and obtain a result which is in good agreement with our first measurement. Using the two measurements, we determine a weighted average slope of $\frac{1}{2}\{ \alpha_0 (7P_{3/2}) + \alpha_2(7P_{3/2}) - \alpha_0 (6S_{1/2}) \} =  4.0389 \: (13)$ MHz/(kV/cm)$^2$.  Using $\alpha_0 (6S_{1/2})$ from Ref.~\cite{AminiG03}, we obtain $ \alpha_0  + \alpha_2 = 8.1776 \, \: (26)$ MHz/(kV/cm)$^2$ for the $7p ^2P_{3/2}$ state. This uncertainty accounts for statistical effects only.  

As we discussed earlier, the frequency difference between the hyperfine components of the Stark spectrum is quantified through the tensor polarizability $\alpha_2$, for which we base our determination on a measurement of the frequency difference between the (5,5) peak and the (4,2) peak.   
	\begin{figure}
	  % Requires \uWesepackage{graphicx}
	  \includegraphics[width=8cm]{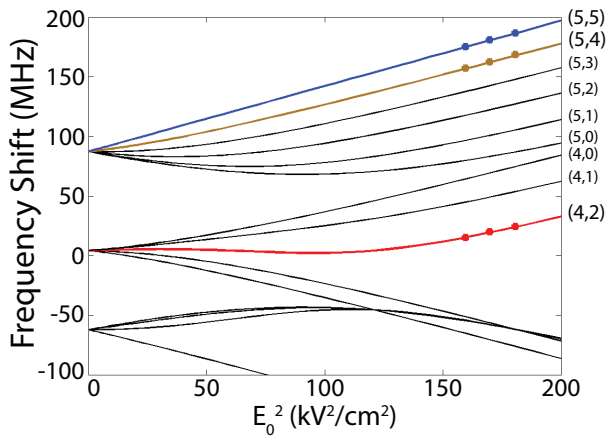}\\ %Was 8cm width
	  \caption{(Color on-line) Frequency $f_{EO}$ for the (5,5) (blue), (5,4) (gold), and (4,2) (red) peaks vs $E_0^2$ for determination of the tensor polarizability $\alpha_2$.  The solid lines are the calculated Stark shifts, less the $\alpha_0$ terms, with $\alpha_2$ adjusted to minimize the deviation with the data. The notation beside each line indicates $(F^{\prime}, \: |m_F^{\prime}| )$ of the $7p ^2P_{3/2}$ state.}
	  \label{fig:Schmieder52fit2}
	\end{figure}
At zero field, the frequency difference between these peaks is the hyperfine splitting $f_{4-5}=83.025~(30)$ MHz~\cite{ArimondoIV77}. We measure this frequency difference by fixing the voltage applied to the field plates (and hence electric field) and adjusting the frequency $f_{EO}$ to bring the Stark laser sideband into resonance with each sublevel. For each of three voltage levels, we repeated the measurement twice. 
In order to determine the value of $\alpha_2$, we fit the diagonalized matrix $Q$ to the data points.  In this case, we fixed the value of $\alpha_0 + \alpha_2$ to the value $8.1776\, \: (26)$ MHz/(kV/cm)$^2$, as discussed above, and used Eqs.~(40a) and (41) from Ref.~\cite{Schmieder72} to generate curves for varying values of $ \alpha_2$. 
The least rms deviation between the calculated and measured frequency difference between the $(F^{\prime}, \: |m_F^{\prime}| ) = (5,5)$ and $(4,2)$ 
Stark-shifted frequencies yields $\alpha_2 = -1.0981 \: (65)$ MHz/(kV/cm)$^2$ $= - 4,413 \: (26) \: a_0^3$.  
In Fig.~\ref{fig:Schmieder52fit2} we show the best fit curves for the Stark-shifted hyperfine peaks versus $E_0^2$, with the scalar component of the Stark shift suppressed. The circles denote our experimental data points. We also show measurements of the $F^{\prime}=5$, $m_F^{\prime}=4$ sublevels in this figure.  We did not use these values in the determination of $\alpha_2$, since the smaller frequency difference between this peak and the (5, 5) gave these values a larger uncertainty and larger error. 
Combining our results for $\alpha_0 + \alpha_2$ and $\alpha_2$ yields a value of $\alpha_0 = 9.2757 \: (70) $ MHz/(kV/cm)$^2$, or $37,277 \: (28) \: a_0^3$.

\section{Measurement errors}\label{sec:errors}
The uncertainties in the polarizabilities that we presented in the previous section include only statistical effects derived from the scatter in the data points from the fitted lines.  In addition, there are other experimental factors, as summarized in Table~\ref{table:sourcesoferror}, that we must consider.  In this section, we discuss these contributions and provide estimates of their magnitudes.

\begin{table}

\begin{tabular}{l c c}
	\hline\hline
	\multicolumn{1}{c}{Source}   & Uncertainty & $\%$ of $\alpha$ \\ \hline
	Field plate spacing          & 4 $\mu$m    & 0.16             \\
	Voltage divider ratio        & 0.005$\%$   & 0.01             \\
	Voltage measurements         & 0.005$\%$   & 0.01            \\
	Error signal line center     & 0.2 MHz     & 0.02             \\
	AOM drive frequency          & 10 kHz      & 0.01             \\
	EOM drive frequency          & 10 kHz      & 0.01             \\
	Beam alignment into chamber  & 0.05 mrad   & 0.01             \\
	                             &             &  \\
	Total systematic uncertainty &             & 0.16             \\
\hline \hline 
\end{tabular} 

\caption{Sources of error, estimates of their uncertainties, and the resulting percentage uncertainty in $\alpha$ resulting from each source.}
\label{table:sourcesoferror}
\end{table}

The largest uncertainty in our measurements is the systematic effect due to determination of the static electric field strength.  These uncertainties derive from the uncertainty in the field plate spacing (including the uncertainty in the measurement of this spacing $d$, as well as any non-uniformity in $d$), the uncertainty in the measurement of the voltage applied to the plates, and edge effects that reach in to the center of the field plates.  We have discussed the first of these in Section~\ref{sec:apparatus}, where we estimate an uncertainty of the plate spacing of 0.08\%.  Since the Stark shift depends on $E_0^2$, the corresponding uncertainty in the polarizabilities is twice as large, or 0.16\%.  
We also described the voltage dividers that we used to measure the voltage applied to the field plates in Section~\ref{sec:apparatus}. This fractional uncertainty of $5\times 10^{-5}$ results in an uncertainty in the polarizabilities of $1\times 10^{-4}$. We also list in Table \ref{table:sourcesoferror} the measurement error of the voltmeter as specified by the manufacturer.

We estimate that the precision with which we can measure the linecenter of each of the Stark shifted lineshapes is $\pm 0.2$ MHz. This is primarily limited by signal asymmetry due to residual amplitude modulation of the Stark beam at 145 Hz. For instance, if the asymmetry of a dispersion-shaped resonance is 15\% of the maximum error signal, as was typical of our measurements, the zero crossing is shifted by $\sim$0.2 MHz, assuming a 6 MHz linewidth of the absorption peak. 
Another limiting factor is dc offsets in the error signal, due to electronics and the overlap from adjacent peaks in the spectrum. We modeled the pulling of the line center due to adjacent peaks and found its effect on the polarizabilities to be less than $2\times10^{-5}$. We estimate that these limiting factors lead to a fractional uncertainty in the polarizabilities of $2 \times 10^{-4}$.

We also considered frequency shifts due to changes in the propagation direction of the laser beam.  Such a change could introduce a Doppler shift in line center of the resonance. Heating effects in the EOM could deflect the beam, for example.  We have projected the Stark beam onto a screen 10 m beyond the EOM, and were unable to observe any such deflection.  We place an upper limit of 0.05 mrad on any such shift. Estimating the Doppler shift to be about 0.7 MHz/mrad, this shift corresponds to an uncertainty of less than 0.05 MHz.  
This limit is consistent with our observed rms residuals of the measured peak positions in Figs.~\ref{fig:p12poscorrStraightfitCombined} and~\ref{fig:p32posStraightfitCombined} of 0.2 MHz.

In order to determine $\alpha_2(7P_{3/2})$ from our measurement of the frequency difference between the (5,5) and (4,2) peaks of the stark spectrum, we used the hyperfine constants $a = 16.605~(6)$ MHz and $b = -0.15~(3)$ MHz from Ref.~\cite{ArimondoIV77}. We consider here the effect of the uncertainty of these hyperfine constants on the uncertainties of the polarizabilities of the $7P_{3/2}$ state. By varying the values of the constants by one standard deviation and running the fitting function again, we can estimate their effect on our values of $\alpha_0(7P_{3/2})$ and $\alpha_2(7P_{3/2})$. This effect is estimated to be 0.21\% for $\alpha_2(7P_{3/2})$ and $<0.02\%$ for $\alpha_0(7P_{3/2})$.

	% $<$ $2 \times 10^{-4}$ except in the case of $\alpha_2(7P_{3/2})$, where changing the value of $a$ by one standard deviation causes a $0.21\%$ change in polarizability. 

For our measurements of the scalar polarizabilities $\alpha_0(7P_{1/2})$ and $\alpha_0(7P_{3/2})$, only the 0.08\% variability in the field plate spacing is significant. These effects contribute a 0.16\% uncertainty. For the tensor polarizability $\alpha_2(7P_{3/2})$, there is an additional 0.21\% error due to the uncertainty in the hyperfine constants. We add these uncertainties in quadrature with the statistical uncertainty stated earlier to obtain the total uncertainty.  For the scalar polarizabilities $\alpha_0(7P_{1/2})$ and $\alpha_0(7P_{3/2})$, this results in an uncertainty in the final result of 0.17\% and 0.18\% respectively.
For the tensor polarizability $\alpha_2(7P_{3/2})$, the statistical uncertainty is the primary contributor to the 0.66\% uncertainty in our result. In the next section, we present our final results for each, and compare with prior experimental and theoretical determinations of these quantities.

\section{Discussion}\label{sec:discussion}

\begin{table}[t]
\begin{tabular}{|l|l|l|l|}
  \hline
    \multicolumn{1}{|c|} {Group}   & $ 7p^2P_{1/2}$        &   \multicolumn{2}{c|}{$ 7p^2P_{3/2}$ }      \\ \hline 
               &  \multicolumn{1}{c|} {$\alpha_0$}    &  \multicolumn{1}{c|} {$\alpha_0$}   & \multicolumn{1}{c|} {$\alpha_2$}            \\  \hline \hline \multicolumn{1}{|l|}{\underline{\emph{experiment}}} & & &  \\
   Khadjavi, {\it et al.}, Ref.~\cite{KhadjaviLH68}    &  &   & $-4.33 \: (17)$    \\
  Khvostenko                                                       &   &   &                        \\
  \hspace{0.2in}and Chaika, Ref.~\cite{KhvostenkoC68}        &   &   & $-3.9  \:  (1) $     \\
 Domelunksen, Ref.~\cite{Domelunksen83}				& $29.5 \: (6) $ & $37.8 \: (8)$ &  $-4.42 \: (12)$ \\
 This work                                         &  $  29.66 \: (5)  $ &  $37.28 \: (7) $ & $ -4.413 \: (29)  $    \\
 & & & \\
\multicolumn{1}{|l|}{\underline{\emph{theory}}} & & &\\
   van Wijngaarden                                            &         &            &   \\
  \hspace{0.2in}and Li, Ref.~\cite{WijngaardenL94}   & $29.4$ &  $36.9$   & $ -4.28 $   \\
  Iskrenova-Tchoukova,  &    &           &              \\
  \hspace{0.2in} {\it et al.}, Ref.~\cite{IskrenovaSS07}   & $  29.89 \: (70)  $ & $ 37.52 \: (75) $ & $ -4.41 \: (17)  $   \\
 \hline
\end{tabular}
\caption{Comparison of the polarizabilities determined in this work to those of prior experimental and theoretical works. All values are in units of $10^3 \: a_0^3$.}
\label{table:ResultComparison}
\end{table}

Our results are in good agreement with, and of higher precision than, previous theoretical and experimental results.  We present a summary of past theoretical and experimental results in Table~\ref{table:ResultComparison}.
Our measurement results for the scalar polarizabilities are
\begin{equation}
\alpha_0(7P_{1/2}) = 29,660 \: (50) \: a_0^3  
\end{equation}
and
\begin{equation}
\alpha_0(7P_{3/2}) = 37,280 \: (70) \: a_0^3  .
\end{equation}
For the tensor polarizability of the $7P_{3/2}$ state, we find 
\begin{equation}
\alpha_2(7P_{3/2}) = -4,413 \: (29) \: a_0^3.  
\end{equation}
Each of these results agrees with past measurements reported in Refs.~\cite{KhadjaviLH68} and \cite{Domelunksen83}.  The measurement result for $\alpha_2$ for the $7P_{3/2}$ in Ref.~\cite{KhvostenkoC68} differs by $\sim 10\%$ from the others, including the present results.  The precision of the present results is much higher than that of the previous reports, due to our use of narrow-band laser sources, Doppler-free resonances, and r.f. modulation techniques.  The theoretical calculations of van Wijngaarden and Li~\cite{WijngaardenL94} and of Iskrenova-Tchoukova {\it et al.}~\cite{IskrenovaSS07} are in good agreement with our results for all three polarizabilities as well.  The former does not report uncertainties.  Our results differ from those of Ref.~\cite{IskrenovaSS07} by typically less than 1\%, while their stated uncertainties are about 2\%.

\section{Conclusion}\label{sec:conclusion}

We have described our experimental determinations of the Stark shift of the $6s \: ^2S_{3/2} \rightarrow 7p \: ^2P_{J} $ transitions for J = $\frac{1}{2}$ and $\frac{3}{2}$ in atomic cesium. Through use of a narrowband, frequency-stabilized diode laser and Doppler-free techniques, the precision of our measurements is higher than that of previous measurements. 
While our polarizability measurements do not yield radial matrix elements directly, the strong agreement between our polarizabilities and those of Ref.~\cite{IskrenovaSS07} do infer that the radial matrix elements calculated in that work are very accurate. 

%Our results provide good confirmation of calculated dipole transition moments presented in Ref.~\cite{IskrenovaSS07}.

This material is based upon work supported by the National Science Foundation under Grant Number PHY-0970041.

\end{document}